\documentstyle[aps,prl,preprint]{revtex}
\begin{document}
\preprint{Submitted to the Physical Review B}
\title
{Universal Behavior of the Spin-Echo Decay Rate in La$_2$CuO$_4$}
\author{\large Andrey V.\ Chubukov}
\address{
Department of Physics, University of Wisconsin-Madison, Madison WI 53706 \\
and P.L. Kapitza Institute for Physical Problems, Moscow, Russia}
\author{\large Subir Sachdev}
\address{Departments of Physics and Applied Physics, P.O. Box 6666, \\
Yale University, New Haven, CT 06511}
\author{\large Alexander Sokol}
\address{
Department of Physics and Materials Research Laboratory, \\
University of Illinois at Urbana-Champaign, Urbana, IL 61801 \\
and L.D.\ Landau Institute for Theoretical Physics, Moscow, Russia}

\maketitle

\begin{abstract}
We present a theoretical expression for the spin-echo decay rate, $1/T_{2G}$,
in the quantum-critical regime of square lattice quantum
antiferromagnets. Our results are
in  good agreement with recent experimental data by Imai
{\em et al.\/} [Phys. Rev. Lett. {\bf 71}, 1254 (1993)] for
$\rm La_2CuO_4$.
\end{abstract}
\narrowtext

\section{INTRODUCTION}
Part of the recent interest in high-$T_c$ superconductivity
has been devoted to the study of low temperature magnetic phases of
undoped and weakly doped antiferromagnetic $La_2CuO_4$ and related compounds.
Neutron scattering measurements
of the correlations length at $T \leq 500K$\cite{Keimer}
have  established that it
agrees very well with the theoretical result
for the low temperature renormalized-classical (RC)
region~\cite{Polyakov,CHN,Hasen}~(we set $k_B =1$)
\begin{equation}
\xi \sim 0.34~\left(\frac{\hbar c}{2 \pi \rho_s}\right)~ \exp(2\pi \rho_s/T)~
\left[1 - \frac{T}{4\pi \rho_s} +...\right].
\label{1}
\end{equation}
 In this region, the
correlation length is determined predominantly
by classical, thermal fluctuations,
 while quantum fluctuations only renormalize the values of input
$T=0$ parameters - the spin stiffness $\rho_s$ and the spin-wave velocity $c$.
The RC value of the correlation length was also successfully
used~\cite{Chakravarty-Orbach} to fit the data on longitudinal
spin-lattice relaxation rate, $1/T_1$, for $400K <T<600K$~\cite{Imai}.

More recently, significant attention has also been devoted to a region of
intermediate temperatures where the temperature becomes larger
than the energy scale associated with  the spin-stiffness at $T=0$.
Under these circumstances,
the system behaves almost as if $\rho_s =0$, i.e. as if it is
at the quantum transition point separating the N\'{e}el and
quantum-disordered states.
This region, where the temperature is the largest energy scale in the problem,
was first
identified in~\cite{CHN} as the quantum-critical (QC) region, and then
studied in detail in~\cite{Subir,Chubukov-Sachdev}.
Clearly, in the QC region, all Bose factors are of order of unity,
and hence quantum and classical fluctuations are equally important.

It seems appropriate at this point to review some of the general ideas
on the RC/QC crossover, and to present the aims of
the theoretical comparisons with experiments. All two-dimensional
magnets with a N\'{e}el ordered ground state are characterized by three
energy scales: $\rho_s$, $T$, and a near-neighbor exchange constant $J$.
It was pointed out in Ref.~\cite{Chubukov-Sachdev} that when $\rho_s / J$ and
$T/J$
are small,  the properties of such
antiferromagnets are characterized by {\em universal} functions of $T/\rho_s$.
The physics is a smooth function of $T/\rho_s$, especially
simple in two limits: the RC limit ($T/\rho_s \rightarrow 0$) and
the QC limit ($T/\rho_s \rightarrow \infty$). While only approximate
$N=\infty$ results have been obtained for the full functional dependence on
$T/\rho_s$, rather precise numerical predictions are available in the
two limits. At intermediate values of $T/\rho_s$, deciding which of
the two limits is appropriate is subject to interpretation, and different
measurements may lead to different conclusions: we shall present
a case for our choices below. A second, important issue is that of
the corrections of order $\rho_s / J$ and $T/J$.
These corrections are in fact
{\em non-universal\/}, depend upon the nature of the lattice cutoff,
and impossible to calculate precisely in an analytical theory.
As theorists, the best we can do is to calculate the universal parts,
compare with the experiments and numerical simulations,
 and then determine whether the
discrepancies can be attributed to these non-universal corrections.
Clearly, $\rho_s /J$ corrections will eventually become important at
high enough temperatures - this  will then impose
a model-dependent upper boundary of the QC region. It turns out however, that
in the temperature range of experimental observations,
 the universal terms agree surprisingly well
with the data indicating that these non-universal corrections are
in fact quite small. Why this should be so is not completely understood,
but we can offer the following plausibility arguments:
({\em i\/}) For small $T/\rho_s$, Hasenfratz and Niedermayer~\cite{Hasen}
have shown that there are no $\rho_s/J$ corrections
to the leading terms in the low temperature expansions of observables;
({\em ii\/})
numerical studies of the dynamical susceptibility, $\chi(k, \omega)$, have
shown~\cite{Sokol3} that in the whole temperature range of experiments
it remains strongly peaked at ($\pi,\pi$), the local
susceptibilities related to NMR and NQR measurements are then given
by momentum integrals confined to $k \ll \Lambda$ where $\Lambda$
is the upper cutoff in momentum
space; ({\em iii\/}) recent numerical analysis of the cutoff-dependent,
two-loop equations of the sigma model~\cite{unpub}, shows almost no cutoff
dependence in the QC results until $T \sim 0.5 c \Lambda$.
We will argue below
that the upper boundary of the universal QC region in the spin-1/2 square
lattice Heisenberg model (above which $T/J$ terms cannot be neglected)
is about $0.6J$ ($\sim 900K$ in $La_2CuO_4$, we use $J=1500K$).
This temperature
approximately coincides with the maximum temperature of
experimental measurements.

We begin by presenting our scenario of where the RC/QC crossover
occurs in the universal  functions of $T/\rho_s$.
Despite the fact that
in Eqn. (\ref{1}) the dimensionless ratio is $T/2\pi \rho_s$,
two of us have recently shown
explicitly~\cite{Chubukov-Sachdev} that the parameter which governs the
crossover behavior between RC and QC regimes is three times larger:
$x = 3T/2 \pi \rho_s$.
The reason for the extra factor of three is the following.
{}From renormalization-group studies, it is known that the
effective spin-stiffness which has to be compared with the temperature, is
$\rho^{eff}_s =2 \pi \rho_s/\widetilde{N}$
where $\widetilde{N}$ is the effective number of
components of the order parameter which contribute to the coupling constant
renormalization~\cite{Polyakov}.  Deep in the RC region,
longitudinal fluctuations are suppressed, and the running
coupling constant diverges at the scale of correlation length
only due to {\it interactions}
 between $N-1$ transverse spin-wave modes (for $O(N)$ systems).
 In this situation, $\widetilde{N} = N-2$, i.e, $\widetilde{N}=1$ for $O(3)$
case.
 On the contrary, in the crossover
region near the quantum phase transition,
all fluctuation modes contribute equally to the correlation length
and we have simply
$\widetilde{N} =N$. For $O(3)$ magnets then $\rho_{eff} = 2 \pi \rho_{s}/3$,
and
therefore $x = T/\rho_{eff} = 3T/2 \pi \rho_s$.
It is the variation of $\rho^{eff}_s$ vs $T/\rho_s$ which eventually
gives rise to $N \rightarrow N-2$ substitution for $\rho_{eff}$
deep in the RC region, as was observed in~\cite{Chubukov-Sachdev}.

For the $S=1/2$ antiferromagnet on a square lattice, both
perturbative~\cite{pert}
and numerical~\cite{Singh} studies yield $\rho_s =0.18J,~c = 1.67Ja$, so that
$x = T/0.38J$. On general grounds, we expect
the crossover between RC and QC
regimes to occur around $x \sim 1$, although not necessarily
at the same $x$ for all observables. The upper boundary
of the universal QC region is, we said, around $0.6J$. Then,
in the undoped  antiferromagnets, the temperature range of the QC
behavior should be $0.4J <T<0.6J$, which for $La_2CuO_4$
corresponds to $600K < T< 900K$ - a range which is accessible to experimental
studies (a wider QC region is expected in the doped cuprates~\cite{comment}).
 The  uniform susceptibility data
in this range~\cite{Johnston,DM} was compared to both
RC~\cite{Hasen,Chubukov-Sachdev} and QC~\cite{Chubukov-Sachdev}
formulas, and very good agreement with the QC result was
found. Furthermore, the measured spin-lattice relaxation rate, $1/T_1$,
was found to be nearly temperature-independent above $700K$,
which is consistent
with the QC behavior; the measured
constant value of  $1/T_1 \sim 2.7 \, \times \,
10^3 sec^{-1}$ is also quantitatively reproduced by the QC theory
of Ref~\cite{Chubukov-Sachdev} which predicts  $1/T_1 \sim (3.2 \pm 0.5) \,
\times
\, 10^3 sec^{-1}$.

Very recently, Imai {\em et al.\/}~\cite{expt2}
reported results on the Gaussian component of the spin-echo
decay rate $1/T_{2G}$ in the temperature range between $450K$ and $900K$.
Much can be learned from their results, but we will need
theoretical predictions for $T_{2G}$ in the QC region, which are computed
in the following section.

\section{$1/T_{2G}$ in the quantum-critical region}

The Gaussian component of the spin-echo decay rate $1/T_{2G}$ is
related to the {\em static} susceptibility,
$\chi ( {\bf q} ) \equiv \chi({\bf q},\omega=0)$,
by~\cite{Slichter}
\begin{equation}
\left(\frac{1}{T_{2G}}\right)^2 = \frac{1}{a^2}~
\int \frac{d^2 q}{4 \pi^2} A^{4}_{\perp} ({\bf q}) ~
(\chi ({\bf q}))^2 - \left[\int \frac{d^2 q}{4 \pi^2}
A^{2}_{\perp} ({\bf q}) ~ \chi ({\bf q})\right]^2
\label{2}
\end{equation}
where $a$ is the interatomic spacing and $A_{\perp} ({\bf q})$
is a formfactor in the direction perpendicular to $CuO_2$ plane,
which early studies~\cite{Monien}
estimated as $A^{2}_{\perp} (\pi,\pi) = 4.85 \cdot 10^{7}K~/sec$.
In the temperature region of interest, $\chi ({\bf q})$
is strongly peaked at ${\bf q} =(\pi,\pi)$.
The second term in (\ref{2}) is nonuniversal in two
dimensions, but it contains one less power of the correlation length
compared to the first term,  and thus can
safely be neglected at low temperatures where
$\xi \gg  a$. To the same accuracy, we have to take the form-factor exactly
at ${\bf q}=(\pi,\pi)$, in which case $(1/T_{2G})^2$  measures simply
the local static $\chi^2$.

Deep in the quantum-critical region,
$1/T_1 \propto T^{\eta}$, $\chi (q) \propto
q^{-2+\eta} f(q/T)$ ~\cite{Subir}, where $f(\infty)$ is a constant, and
 $\eta = 0.028$~\cite{Holm,Ma}
is the critical exponent for the spin correlations at criticality.
 A straightforward analysis
then yields~\cite{Sokol} $1/T_{2G} \propto T^{-1 +\eta}$.
One may combine this with previous results on $1/T_1$, and the complete
scaling forms (including prefactors) obeyed by the staggered
susceptibility~\cite{Chubukov-Sachdev} to obtain
\begin{equation}
\frac{T T_1}{T_{2G}} = \left(\frac{A_{\perp}(\pi,\pi)}
{A_{\parallel}(\pi,\pi)}\right)^2
\frac{\hbar c}{a} {\cal R}
\end{equation}
where ${\cal R}$ is a {\em universal\/} number, computable
in the $1/N$ expansion. Here $A_{\parallel} (\pi,\pi)$
is the in-plane hyperfine formfactor
which appears in the $1/T_1$ measurements of Ref.\cite{Imai}.
The ratio $T T_1/T_{2G}$ is indeed found to be temperature independent
in the data of Imai {\em et al.\/}~\cite{expt2}. However
for {\em quantitative}
 comparisons of $1/T_{2G}$ data with the theory, and for the estimate
of the correlation length, Imai {\em et al.\/} used
a RC expression
for $\chi(q)$~\cite{CHN}
 modified by finite-temperature corrections in a manner first
discussed by Shenker and Tobochnik~\cite{Shenker}.
Below we take an alternative
approach and compare the experimental data to
 the QC formula for $1/T_{2G}$ which we derive here.
We will see that the agreement between our theory and experiment is
rather good, and thus confirm the original conclusion of~\cite{Sokol,expt2}
 that the data on $1/T_{2G}$ favor QC behavior at intermediate
temperatures.

As input for our calculations, we need the expression for the static
susceptibility near the antiferromagnetic wave vector. In the QC
region, the only energy scale is the temperature, and the scaling function
for the susceptibility depends only on ${\bar q} = \hbar c q/T$. Using
the results of Ref~\cite{Chubukov-Sachdev}, we then obtain
\begin{equation}
\chi (q) = \frac{N^{2}_0}{\rho_s}~\left(\frac{N T}{2 \pi
\rho_s}\right)^{\eta}~\left(\frac{\hbar c}{T}\right)^2 \Phi ({\bar q}),
\label{3}
\end{equation}
where $\Phi ({\bar q})$ is a universal function  given by
\begin{equation}
\Phi ({\bar q}) = \frac{Z}{{\bar q}^2 + m^2 +
\Sigma ({\bar q})}.
\label{4}
\end{equation}
Here $Z$ is a rescaling factor,
$m$ is proportional (but not exactly equal) to inverse correlation
length,
and $\Sigma ({\bar q})$ is a self-energy given by
\begin{equation}
\Sigma ({\bar q}) = T\sum_{n}\int \frac{d^2 {\bar k}}{4 \pi^2}~\frac{G
({\bar k}+{\bar q},{\bar \omega}_n) -G({\bar k}, {\bar \omega}_n)}{\Pi
({\bar k}, {\bar \omega}_n)},
\label{5}
\end{equation}
where ${\bar \omega}_n = \hbar \omega_n /T, ~
G({\bar k}, {\bar \omega}_n) = 1/({\bar k}^2 + {\bar \omega}^{2}_n + m^2)$,
and $\Pi ({\bar k}, {\bar \omega}_n)$ is a polarization operator.

Below we calculate $1/T_{2G}$ in the two leading orders in $1/N$ expansion for
$O(N)$ magnets. The physical case will be considered at the end
by setting $N=3$.
To first order in $1/N$ we have~\cite{Chubukov-Sachdev}
\begin{eqnarray}
Z &=&  1 + \eta \log{\frac{\Lambda \pi}{ 8 T}}, \nonumber \\
m^2 &=&  \Theta^2 \left(1 +
\eta \log{\frac{\Lambda}{T}} + \frac{0.231}{N}\right), \nonumber \\
\Sigma ({\bar q}) &=&  {\bar q}^2 \left(\eta \log{\frac{\Lambda}{T}}
\right) + {\bar \Sigma} ({\bar q}).
\label{6}
\end{eqnarray}
Here $\Theta = 2 \log[(\sqrt{5} +1)/2]$,
$\Lambda$ is an upper cutoff and $ {\bar \Sigma} ({\bar q}) \propto 1/N$
 stands for the regular part of the self-energy term.
Substituting Eqn. (\ref{6}) into
Eqn. (\ref{4}), we observe that all
$\Lambda$-dependent terms disappear as they
should, so that the scaling function for $\chi ({\bar q})$ is universal.
Performing then the momentum integration in Eqn. (\ref{2})
and numerically evaluating
the contribution from ${\bar \Sigma} ({\bar q})$,
 we  obtain  after some algebra
\begin{eqnarray}
\frac{1}{T_{2G}} &=&
\frac{A^{2}_{\pi}}{a\sqrt{4\pi}\Theta}~
{}~\frac{N^{2}_0}{\rho_s}~\left(\frac{N T}{2 \pi \rho_s}\right)^{\eta}~
\frac{\hbar c}{T}~\left(1+ \frac{0.22}{N}\right)\nonumber \\
&&\left[1 + {\cal O}\left(\frac{2 \pi \rho_s}{N T}\right)^{1/\nu}\right],
\label{7}
\end{eqnarray}
where $\nu \sim 0.7$ is the critical exponent for correlation length.
It is then convenient to  reexpress the result for $1/T_{2G}$
in terms of the actual correlation length
defined from the exponential ($e^{-r/\xi}$) decay of the spin-spin correlation
function, or, equivalently, from the pole of the static structure factor on
the imaginary $q$ axis. From the large $N$ theory of
Ref~\cite{Chubukov-Sachdev},
deep in the QC region we have
\begin{equation}
\xi^{-1} (T) = \frac{T}{\hbar c}~ \Theta \left( 1 +
\frac{0.237}{N}\right)
{}~\left[1 + {\cal O}\left(\frac{2\pi \rho_s}{N T}\right)^{1/\nu}\right].
\label{8}
\end{equation}
Using Eqn. (\ref{8}), we can rewrite  Eqn. (\ref{7}) as
\begin{eqnarray}
\frac{1}{T_{2G}} &=& \frac{A^{2}_{\pi}}{\sqrt{4\pi}}~
{}~\frac{N^{2}_0}{\rho_s}~\left(\frac{N T}{2 \pi \rho_s}\right)^{\eta}~
 \frac{\xi}{a}~\left (1 + \frac{0.46}{N}\right) \nonumber \\
&&~\left[1 + {\cal O}\left(\frac{1}{N}~\frac{2 \pi \rho_s}{N T}\right)\right].
\label{9}
\end{eqnarray}
The advantage of using Eqn. (\ref{9}) is in the form of the
correction term which now has an extra factor of $1/N$; this is because at
 $N=\infty$ all  corrections related to a deviation from pure criticality
are already absorbed into the correlation length.
We will assume that the remaining
corrections are small and neglect them below.

We now use the values of $N_0,~\rho_s$ and $c$ for
$S=1/2$ Heisenberg antiferromagnet, the same value of the form-factor as
in~\cite{Monien,Sokol}, and rewrite the result for $N=3$ as
\begin{equation}
\frac{1}{T_{2G}} \approx 0.546~\frac{\xi}{a} ~\times 10^4 ~sec^{-1}.
\label{10}
\end{equation}

Finally for the ratio $T_1 T/T_{2G}$ deep in the QC region,
we obtain using previous
theoretical
result~\cite{Chubukov-Sachdev} for $1/T_1$
\begin{equation}
\frac{T_1 T}{T_{2G}}~ \approx  3.36 \times 10^3 K
\label{11}
\end{equation}
Note however, that $1/T_1$ is itself of the order $1/N$, and $1/N$
corrections to $1/T_1$ have not been calculated.

\section{DISCUSSION}

We now turn to a comparison with the experimental results
of Imai {\em et al.\/}~\cite{expt2}.
We first use Eqn. (\ref{10}) and infer from the data the values
 of the correlation length. Fig.\ref{fig1} presents our theoretical result
obtained with no adjustable parameters
together with neutron-scattering data available
 at $T<560K$~\cite{Keimer} and the data of numerical
simulations~\cite{DM,Sokol3}.
 We see that above $700K$ (where $1/T_1$ levels off to a constant
value expected in the QC regime),
 the data inferred from the $T_{2G}$ measurements using the
QC formula
are in  good agreement with  numerical
results. At lower temperatures, the actual correlation length increases faster
than in
Eqn. (\ref{10}), which is a clear signature of a crossover into the
RC regime.

Next, we compare the experimental data
directly with our Eqn. (\ref{7}) for $1/T_{2G} (T)$. The comparison requires
some caution  because the
${\cal O}(0.38J/T)$ corrections in Eqn. (\ref{7}) are
not expected to be small. However, previous studies of $1/T_1$  have
shown~\cite{Chubukov-Sachdev,Sokol2} that between
$700K$ and $900K$, the experimental
 data are surprisingly well described by the pure QC formula with
no temperature-dependent corrections. We calculated $1/T_{2G}$ in the same way,
and again
found  surprisingly good agreement with the experimental data above $700K$ (see
Fig.\ref{fig2}). At present we have no explanation why $\rho_s/T$ corrections
to $T_1$ and $T_{2G}$ are small at $T \sim 0.5J$; it is however difficult to be
more quantitative without calculating $1/N$ corrections to subleading term,
 which has not been done.
Note also that we deliberately do not compare the slope of
 $1/T_{2G} (T)$ with our QC formula. The reason for
{\em not} doing so comes from the results of numerical studies of $1/T_{2G}$
 in a $S=1/2$ Heisenberg antiferromagnet~\cite{Sokol3}. The numerical results
 are presented in Fig.\ref{fig3}. The two curves in Fig.~\ref{fig3} are
the slopes of $T_{2G}$ vs the temperature,
obtained using the full Eqn. (\ref{2}) (these data are consistent with
 experiment for all temperatures measured),
and its truncated version without nonuniversal corrections due to
the second term  and to the momentum dependence of the
form-factor: $ {\widetilde T}_2 = a A^{-2}_{\pi} (\int d^2 q
{}~\chi^{2}(q)/4\pi^2 )^{-1/2}$.
We recall that these nonuniversal corrections can be neglected only
if the correlation length well
exceeds the interatomic spacing.
 We see that in the region of spin-echo decay measurements,
$T_{2G}$ and ${\widetilde T}_2$ are quite close to each other,
 so that the absolute
value of $T_{2G}$ (and hence $\xi$) inferred from the experimental
data should be consistent with the  long-wavelength description.
At the same time, the slopes of $T_{2G} (T)$ and ${\widetilde T}_2 (T)$
differ already by the factor of 1.7 at $900K$
 which means that it is dangerous to compare the experimentally measured
slope of $T_{2G}$ with the theoretical formula.

Finally, our theoretical
result for $T_1 T/T_{2G}$, Eqn. (\ref{11}), also agrees satisfactorily
with the
experimental value $T_1 T/T_{2G} \sim 4.3~\times 10^{3}K$. The difference is
chiefly due to the theoretical result
for $1/T_1$, which is a bit larger than in the
experiments~\cite{Chubukov-Sachdev}. Note also that the experimental
values of $T_1 T/T_{2G}$ remain
temperature independent even at smaller temperatures, where the correlation
length already fits the RC formula.
At the same time, deep in the RC region, one has
$T_1 T/T_{2G} \propto T^{1/2}$~\cite{CHN,Sokol}. This disagreement is
not surprising as the RC result for $T_1 T/T_{2G}$ assumes a
temperature independent uniform susceptibility $\chi_u$~\cite{CHN}.
Numerical studies however indicate~\cite{DM} that $\chi_u$
does not saturate until very low $T \leq
0.2J \sim 300K$.
It is therefore likely that in the temperature range of
experimental comparisons here,
the highly nontrivial downturn renormalization
of the spin-wave velocity, which leads to $c(T) \sim \sqrt{T}$ at $T
\rightarrow 0$~\cite{CHN}, does not occur and
one has $T_1 T/T_{2G} = const$ even when $\xi$ is given by Eq.(\ref{1}).

We now address the issue of whether it is possible to
extend the QC behavior above $0.6J$. This issue is probably
irrelevant for experimental
studies, as no experiments have been done above $900K$,
but it is nevertheless
important for the interpretation of numerical data.
In particular, it has been recently  shown
that the numerical data for the correlation length
at $0.6J<T<J$ can be fitted by either by a modified
RC expression~\cite{Birgeneau,Chakravarty}, or an inverse linear
dependence as in the QC regime~\cite{Chubukov-Sachdev,Sokol3}.
However, as  has already been noted~\cite{Chubukov-Sachdev,Sokol3},
the slope of the inverse linear
dependence in the QC fit was nearly twice as large as in Eqn.
(\ref{8}). Our point
of view  is that above  $ T \sim 0.6J$,
a mean-field description, similar in spirit to the
$N=\infty$ approach is possible,
but as we discussed in the Introduction,
it should definitely include nonuniversal $T/J$ terms.
There are at least two sets of data which support the above conjecture.
The first set are the data for the
uniform susceptibility~\cite{Johnston,DM,series} which clearly indicate
that above $T \sim 0.6J$, the susceptibility tends to approach
a broad maximum produced by
short-wavelength  fluctuations. The second set of data, shown in
Fig.\ref{fig3},
 are the numerical results for $T_{2G}$~\cite{Sokol3}.  We see that
above $0.6J$, not only the slopes but also the absolute values of
 $T_{2G}$, ${\widetilde T}_2$ begin to
differ substantially and their ratio reaches a value of
$\sim 1.7$ at $T=J$. A similar nonuniversal behavior is likely to
hold for the correlation length above $0.6J$,
 though we cannot also exclude
the possibility that the RC formula for the correlation
length extends to higher temperatures than for other observables. The latter
is however  unlikely in view of present results and
recent numerical results for doped
antiferromagnets~\cite{Sokol3}.

To conclude, in this paper we have
presented the theoretical expression for the
spin-echo decay rate in the QC region of 2D
antiferromagnets. We compared our QC result with the experimental
data of Imai {\em et al.\/}~\cite{expt2} and found good quantitative
agreement in the temperature range
between $700K<T<900K$. The temperature dependence of the correlation length
inferred
from the $T_{2G}$ data is in good agreement with neutron-scattering
and numerical
data. We have also argued that above $ T \sim 0.6J$,
lattice effects are relevant
and the use of the universal low-temperature expressions for observables is
unlikely to be justified.

The research was supported by the NSF Grant No. DMR92-24290 (S.S) and
by the NSF Grant No. DMR89-20538
through the Materials Research Laboratory at the University of Illinois
at Urbana-Champaign (A.S.).
We are pleased to thank R.J. Birgeneau, S. Chakravarty, T. Imai,
D. Pines and  C.P. Slichter
for useful discussions and communications, and R.R.P.
Singh and R.L. Glenister for the collaboration in the numerical series
expansion  analysis.

\begin{figure}
\caption{The correlation length versus the temperature.
Solid circles are the data inferred from the experiments on $1/T_{2G}$~
\protect\cite{expt2} using the QC formula,
Eq. (\protect\ref{10}). Diamonds are the
neutron-scattering data of Ref.\protect\cite{Keimer}. The
line represents the results of
numerical studies~\protect\cite{DM,series,Sokol3} at $J=1500K$.}
\label{fig1}
\end{figure}

\begin{figure}
\caption{Experimental and theoretical results for the
spin-echo decay rate $1/T_{2G}$.
The solid circles are the experimental data of
Imai {\em et al.\/}\protect\cite{expt2}.
The asymptotic QC result,
Eq. (\protect\ref{7}), is shown as solid line
in the QC region, $T>600K$, where the agreement
with the experiment is expected.}
\label{fig2}
\end{figure}

\begin{figure}
\caption{The series expansion results~\protect\cite{Sokol3} for the
spin-echo decay rates $T_{2G}$, calculated
using Eq. (\protect\ref{2}) (solid line),
and  ${\widetilde T}_2$, calculated
without the nonuniversal second term in (\protect\ref{2}) and
without the momentum dependence of the prefactor (dotted line).}
\label{fig3}
\end{figure}

\end{document}